\documentclass[aps,twocolumn,prl,showpacs,floatfix,nofootinbib]{revtex4-1}
\usepackage{amsmath}
\usepackage{graphicx}
\usepackage{color}
\usepackage{dcolumn}
\usepackage{bm}

\def\beq{\begin{eqnarray}}
\def\eeq{\end{eqnarray}}

\begin{document}

\title{One cannot hear the density of a drum \\
(and further aspects of isospectrality)}

\author{Paolo Amore}
\affiliation{Facultad de Ciencias, CUICBAS, Universidad de Colima, Bernal
D\'{\i}az del Castillo 340, Colima, Colima, Mexico}

\begin{abstract}
It is well known that certain pairs of planar domains have the same spectra of
the Laplacian operator. We prove that these domains are still isospectral for 
a wider class of physical problems, including the cases of heterogeneous drums 
and of quantum billiards in an external field. In particular we show that the isospectrality is 
preserved when the density or the potential are symmetric under reflections 
along the folding lines of the domain. These results are also confirmed numerically
using the finite difference method: we find that the pairs of numerical matrices obtained
in the discretization are exactly isospectral up to machine precision.

\end{abstract}

\maketitle



In an important and influential paper, Kac proposed an interesting problem, summarized
in the title of the paper itself: "can one hear the shape of a drum?"~\cite{Kac66}.
From a mathematical point of view, the spectrum of a drum corresponds to the set
of eigenvalues of the negative Laplacian on a given planar domain, where the solutions
vanish at the border (Dirichlet boundary conditions). Therefore, Kac's question
can be rephrased as "are there nonisometric planar domains where the Laplacian 
has the same spectrum?". A partial answer to the question comes from Weyl's law:
although in most cases, one does not know the spectrum of a given domain exactly, 
the asymptotic behavior of the eigenvalues is related to the geometrical properties 
of the domain (area, perimeter, \dots). As a result it is possible to distinguish drums with 
different area and perimeter just by hearing their sound. This result however does
not exclude the existence of nonisometric isospectral domains of equal area and perimeter.

\begin{figure}
\begin{center}
\bigskip\bigskip\bigskip
\includegraphics[width=5.5cm]{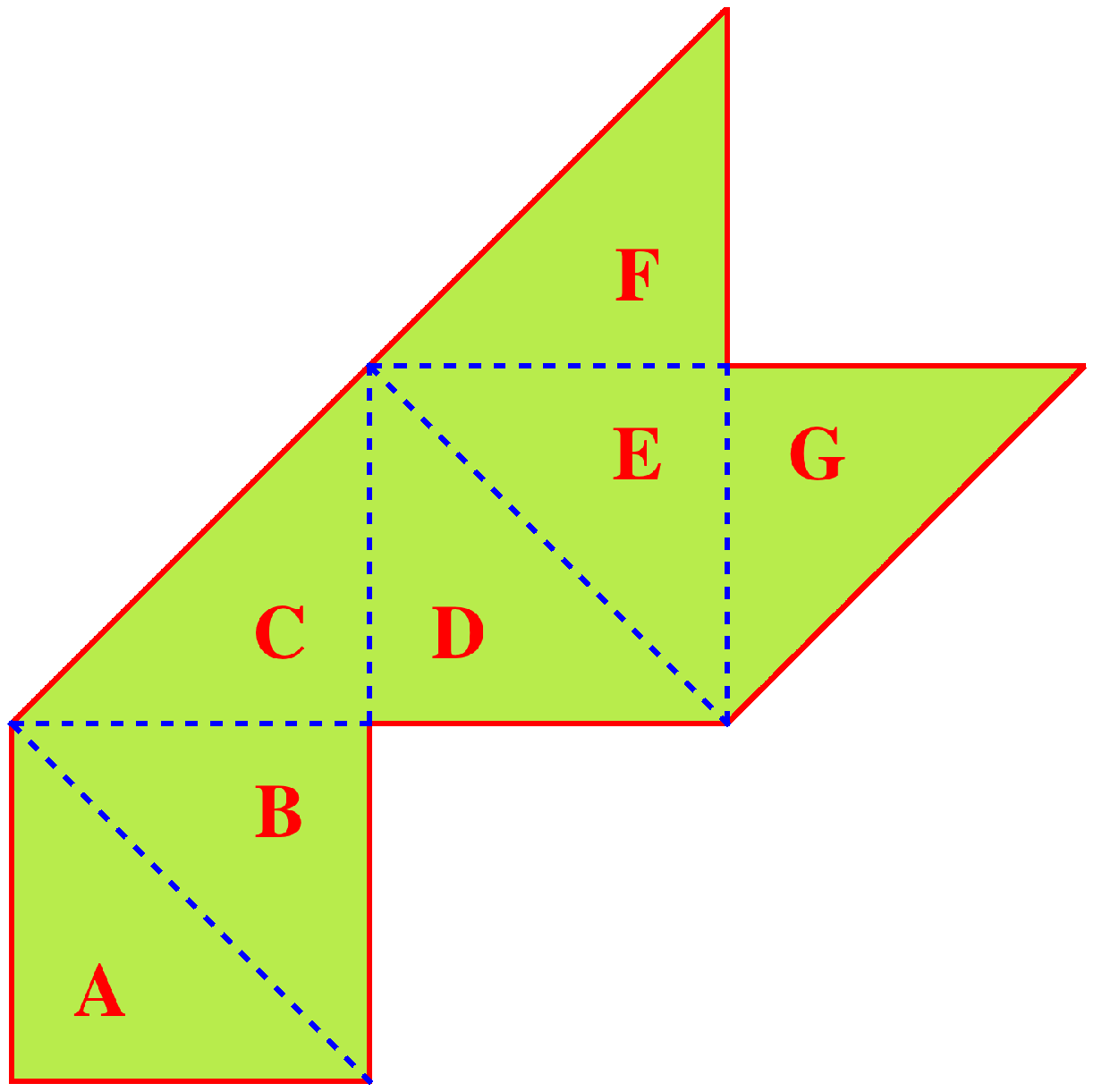} 
\includegraphics[width=5.5cm]{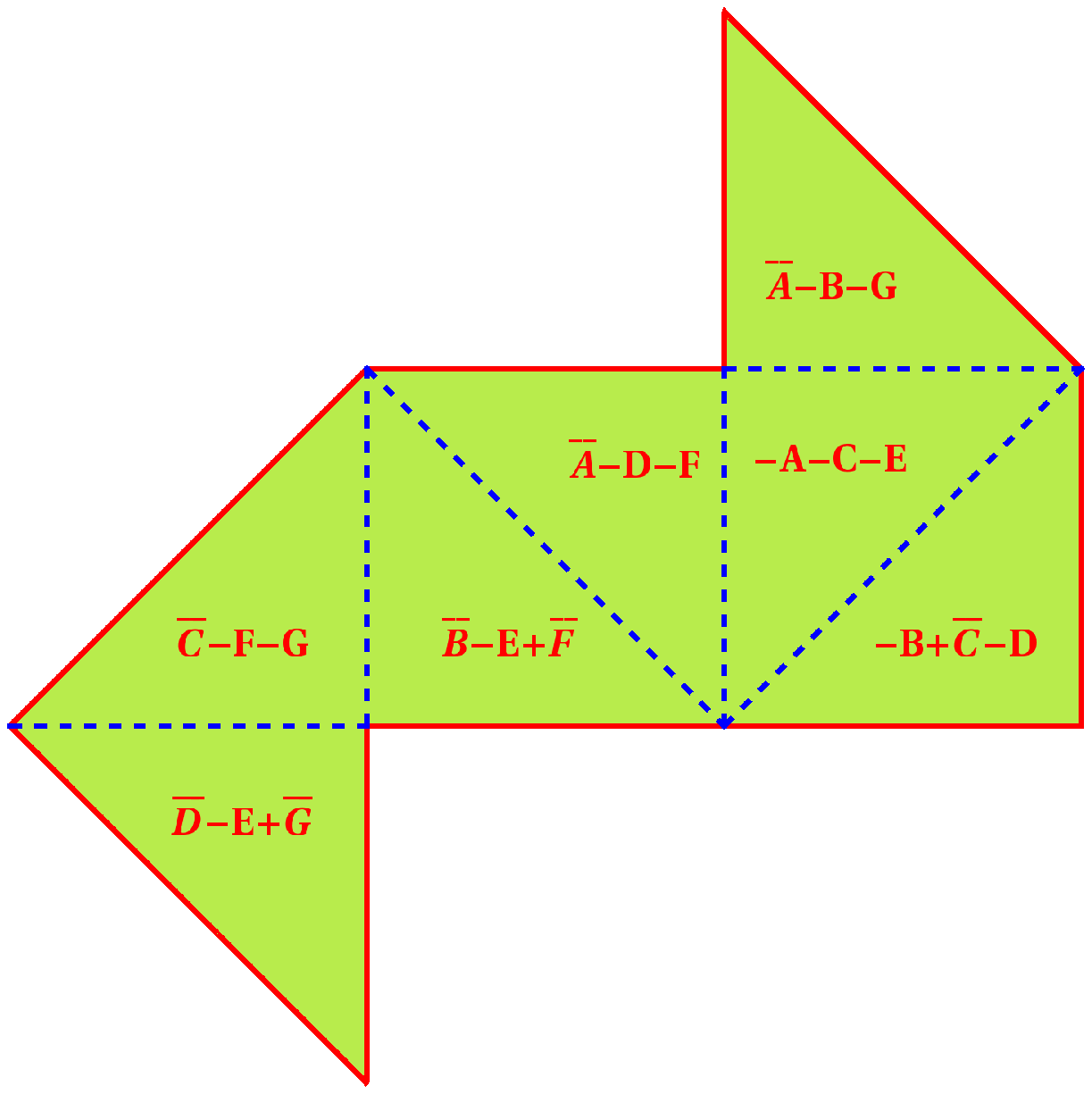} 
\caption{(color online) GWW isospectral drums.}
\label{Fig_2}
\end{center}
\end{figure}

In 1992, twenty five years after the publication of Ref.~\cite{Kac66}, Gordon, Webb 
and Wolpert \cite{GWW92, GWW92b} found a first example of a pair of nonisometric 
planar domains with the same Laplace spectrum (see Fig.~\ref{Fig_2}) using a theorem
by Sunada~\cite{Sunada85}. B\'erard has given a simple proof of the isospectrality constructing a map which
takes an eigenfunction in one domain and maps it into an eigenfunction of the second 
domain \cite{Berard92,Berard93}. Buser et al. \cite{Buser94} have used this "transplantation" 
approach to obtain a large number of isospectral planar domains, while Chapman has visualized 
this result in terms of "paper--folding"~\cite{Chap95}. A discussion of the transplantation
method is also found in \cite{Okada01}. Isospectral domains with fractal border have
been studied by Sleeman and Hua \cite{Sleeman00}.

The isospectrality of these domains has later been verified both 
numerically~\cite{WSM95, Driscoll97} and experimentally using microwave 
cavities \cite{SK94, Even99, Dhar03}. 

More recently, isospectral electronic nanostructures of shapes 
similar to those of Fig.~\ref{Fig_2} have been built by Moon et al~\cite{Moon08}. 
The extra degree of freedom provided by the isospectrality has been used to extract 
the quantum phase of the electron wave functions.
The reader interested in a detailed account of the present state of the research in this area 
should refer to the recent review by Giraud and Thas \cite{GT10}.

In this paper we want to show that it is possible to generalize the results of 
Ref.~\cite{GWW92,GWW92b,Berard92,Berard93} to a wider class of physical problems, such as the 
case of heterogeneous drums or of quantum billiards in an external field.
We will prove that, under certain conditions, the pairs of isospectral domains of the Laplacian 
remain isospectral even in these cases. These results may be summarized saying that one cannot 
hear the shape of an inhomogeneous drum, nor distinguish two quantum billiards in an external 
field uniquely by their spectrum.

A related but different problem has been studied by Gottlieb \cite{Gottlieb04,Gottlieb06} and by
Knowles and McCarthy \cite{KMC04}, who have found examples of materially isospectral congruent 
membranes, i.e. isospectral membranes with the same shape but different densities.  
In particular, the authors of \cite{KMC04} have used a conformal transformation 
on the domains of Fig.~\ref{Fig_2}, obtaining a pair of inhomogeneous isospectral membranes
of circular shape. Holmgren et al. \cite{Capi06} have analysed the problem of hearing the 
composition of an inhomogeneous drum using  tools of asymptotic linear algebra on the 
associated numerical problem.

\textit{Isospectrality---} 
B\'erard has proved that it is possible to map an eigenfunction of one of the domains of Fig.~\ref{Fig_2}
into an eigenfunction of the other domain. Both domains in the figure are 
made of seven building blocks, which are triangles of angles $(45^0, 45^0, 90^0)$. The triangles 
of the first domain are labeled as shown in the figure.

Assuming that an eigenfunction of the first domain is known, the linear combinations 
shown in the second domain of Fig.~\ref{Fig_2} are also solutions of the Laplacian in 
each triangle. Here the notation $\bar{A}$ means that the solution in $A$ is reflected 
with respect to the dashed line. It is easy to see that the function obtained with these
linear combinations and its gradient are everywhere continuous inside the domain and 
that it vanishes at the border. Therefore this function is an eigenfunction of the second
domain with the same eigenvalue.

It is important to notice that the building blocks may be classified into two classes,
where the blocks belonging to the same class are related by an even number of reflections 
along the dashed lines: $\left\{ A,C,E\right\}$ and $\left\{B,D,F,G\right\}$. 
For instance, if we consider the first domain in the figure and we set the origin in the 
upper vertex of the triangle A, we see that a function $f(x,y)$ defined on A transforms 
under reflection along the dashed line into a function $f(y,x)$ on B;
a further reflection along the horizontal dashed line, transforms this function into $f(y,-x)$, 
which can be obtained from the first one with a simple rotation.

One may generate each of the two isospectral domains 
of Fig.~\ref{Fig_2} starting with a single building block with repeated reflections along the dashed lines. 
Notice that the linear combinations in the second domain of Fig.~\ref{Fig_2} only mix functions belonging 
to the same class (observe that under reflection a function changes class).

Consider now the eigenvalue equation
\beq
\hat{H} \psi_n = E_n \psi_n \nonumber 
\eeq
over the first domain of Fig.~\ref{Fig_2}. Here $\hat{H}$ is  a Hermitian operator, which contains the Laplacian
and with an explicit dependence on the coordinates. We assume that we know an eigenfunction of $\hat{H}$ 
and we want to see under what conditions the linear combinations in the second domain of Fig.~\ref{Fig_2} 
provide an eigenfunction of $\hat{H}$. 

In the case of a homogeneous drum, the reflection of the eigenfunction along a dashed line 
is still an eigenfunction of the Laplacian, since this operator commutes with the reflections; 
in the present case, however, because of the explicit dependence on the coordinates, the operator 
does not commute with the reflection and therefore the reflection of the function on $A$, $\bar{A}$, 
is not in general an eigenfunction of $\hat{H}$. 
However, this problem is solved if the operator $\hat{H}$ in each building block is also obtained from the 
operators in the neighboring blocks through a reflection along the dashed line separating
the two blocks.

In this way, the linear combinations appearing in the second domain Fig.~\ref{Fig_2} are once again eigenfunctions
of the operator; since the function obtained with these linear combinations and its gradient are everywhere continuous 
in the domain and the function vanishes on the border, it is an eigenfunction of $\hat{H}$ over the second  domain. Therefore,
{\sl the domains are isospectral}.

It is useful to discuss two physical examples of isospectral problems of this kind.
We consider first the case of an inhomogeneous drum: its vibrations are described 
by the eigensolutions of the Helmholtz equation 
\beq
\left(- \Delta \right) \psi_n(x,y) = E_n \Sigma(x,y) \psi_n(x,y) 
\label{Helmholtz_1}
\eeq
where $\Sigma(x,y)>0$ is the density of the membrane, $(x,y) \in \Omega$, a 
domain in the plane (we also assume Dirichlet boundary conditions on the border, 
$\partial\Omega$, $\psi_n(x,y)|_{\partial{\Omega}} = 0$). 
It is convenient to convert Eq.(\ref{Helmholtz_1}) to
\beq
\left[ \frac{1}{\sqrt{\Sigma}}  \left(- \Delta \right) \frac{1}{\sqrt{\Sigma}} \right] 
\phi_n(x,y) = E_n \phi_n(x,y) 
\label{Helmholtz_2}
\eeq
where $\hat{H} \equiv \frac{1}{\sqrt{\Sigma}}  \left(- \Delta \right) \frac{1}{\sqrt{\Sigma}}$ 
is a Hermitian operator and $\phi_n(x,y) \equiv \sqrt{\Sigma} \psi_n(x,y)$~\cite{Amore10b} 
(notice that Eqs.(\ref{Helmholtz_1}) and (\ref{Helmholtz_2}) have the same spectrum).

\begin{figure}
\begin{center}
\bigskip\bigskip\bigskip
\includegraphics[width=4cm]{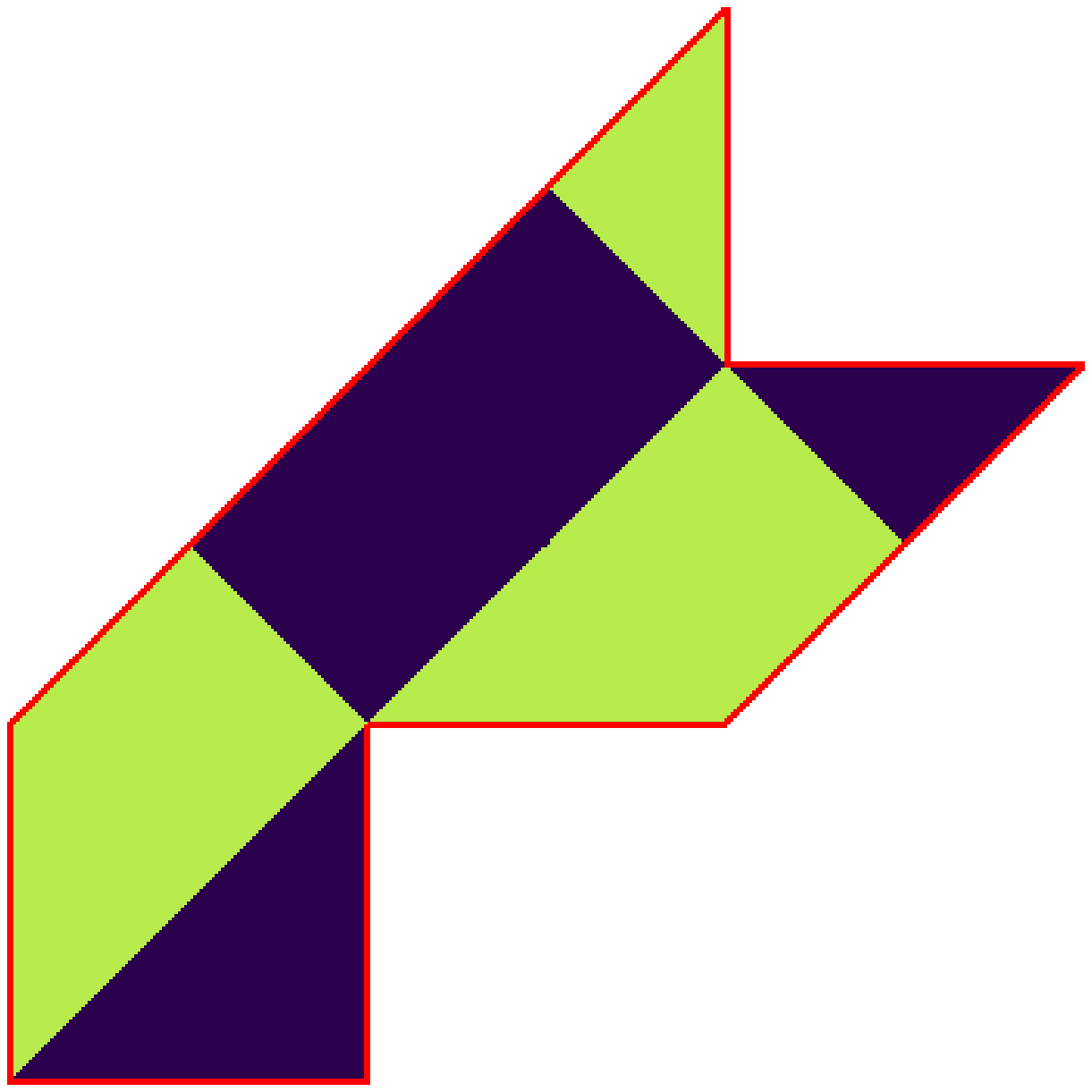}
\includegraphics[width=4cm]{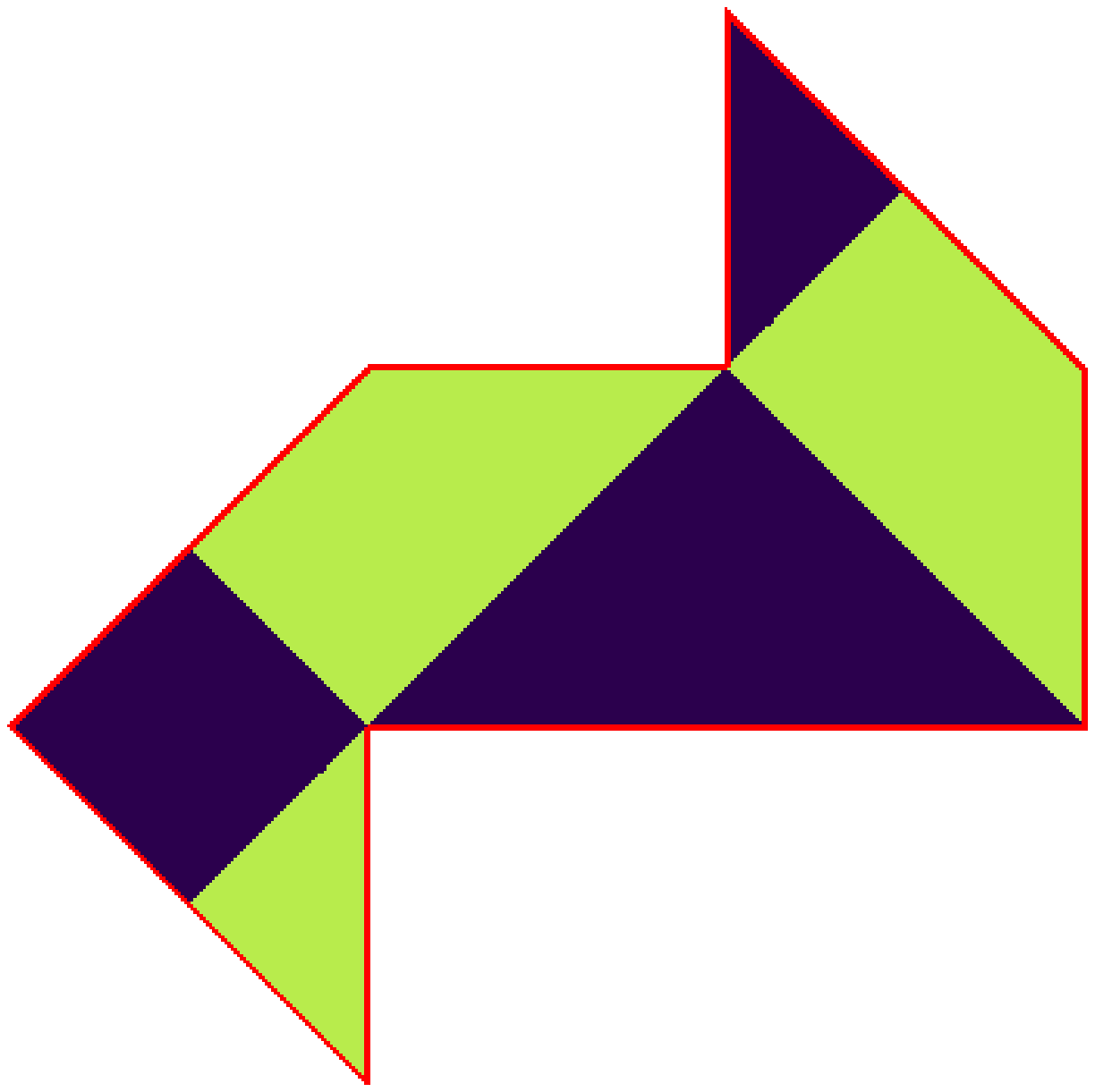} \\
\caption{(color online) Inhomogeneous isospectral drums. The regions with larger density are 
darker.}
\label{Fig_4}
\end{center}
\end{figure}

According to our previous discussion, the two domains will be isospectral if the density $\Sigma$ 
in any of the building blocks that compose the domains is the reflection of the density on a 
neighbouring block along the dashed line separating the two. 
Notice that the two heterogeneous domains have clearly the same mass (we call $\Omega_A$ and $\Omega_B$ the
domains on the left and the right of Fig.\ref{Fig_4} respectively and $\Sigma_A$ and $\Sigma_B$ their
densities) $M = \int_{\Omega_A} dxdy \ \Sigma_A(x,y) = \int_{\Omega_B} dxdy \ \Sigma_B(x,y)$ and therefore
their spectrum has the same asymptotic behavior, provided by Weyl's law, 
$E_n = 4 \pi n/M$ ($n \rightarrow \infty$) \cite{Amore10a}\footnote{Observe that Weyl's 
law allows one to distinguish between inhomogeneous drums of different mass just listening 
to their high frequency sound.}.

Fig.~\ref{Fig_4} displays a pair of inhomogeneous isospectral drums with a piecewise constant 
density (the lighter and darker colors in the figures correspond to two different 
densities $\Sigma_1$ and $\Sigma_2$).
More elaborate examples, with a continuously varying density inside a given block can be 
easily obtained. Also, one could consider more general shapes of the domains, as those 
discussed in \cite{Buser94,Chap95,Sleeman00}.

As a second example of isospectral problem we consider a quantum particle
confined in a finite region under the action of an external force 
(in absence of a force, the operator reduces to the Laplacian, for which 
the isospectrality has already been proved).
Therefore we are interested in the spectrum of the single particle Hamiltonian 
$\hat{H} = \left[-\frac{\hbar^2}{2m } \Delta + V(x,y) \right]$ 
on each of the two domains  of Fig.~\ref{Fig_2}.

\begin{figure}
\begin{center}
\bigskip\bigskip\bigskip
\includegraphics[width=4cm]{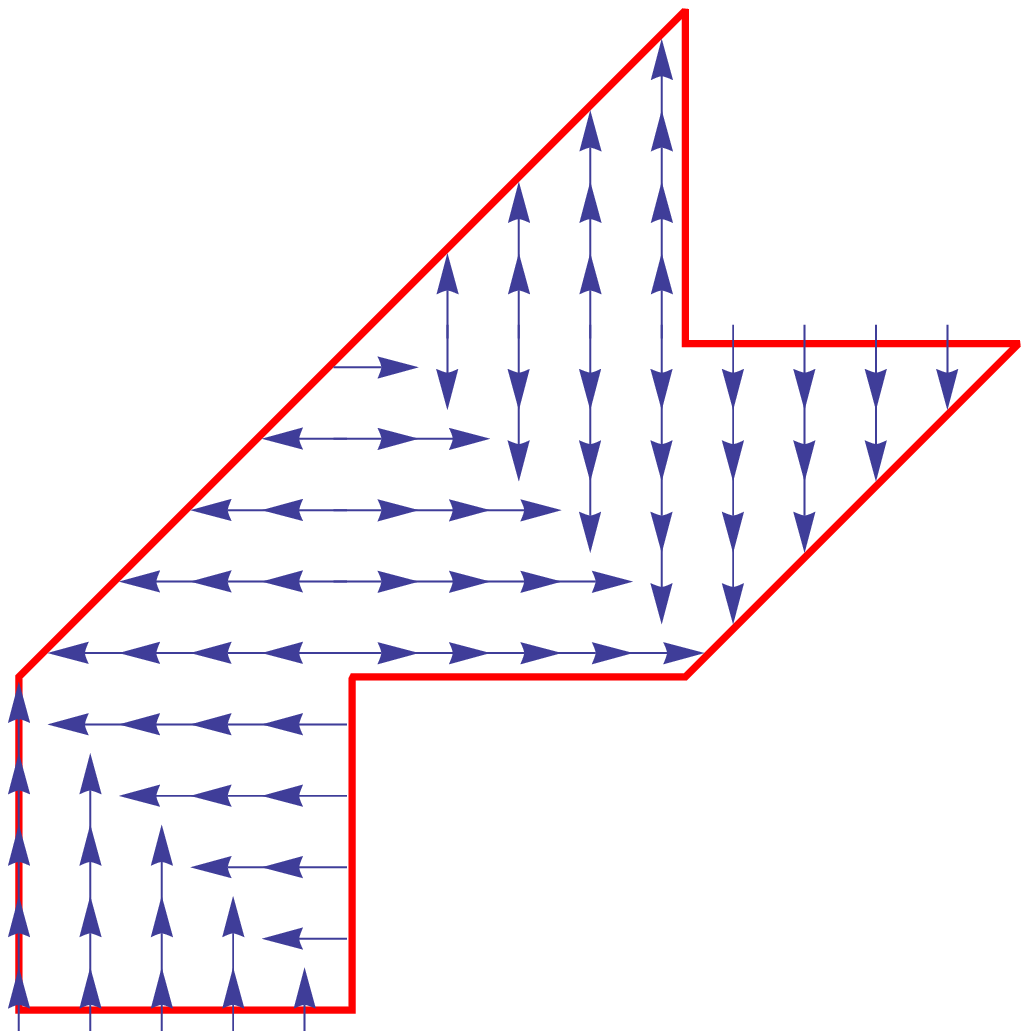} 
\includegraphics[width=4cm]{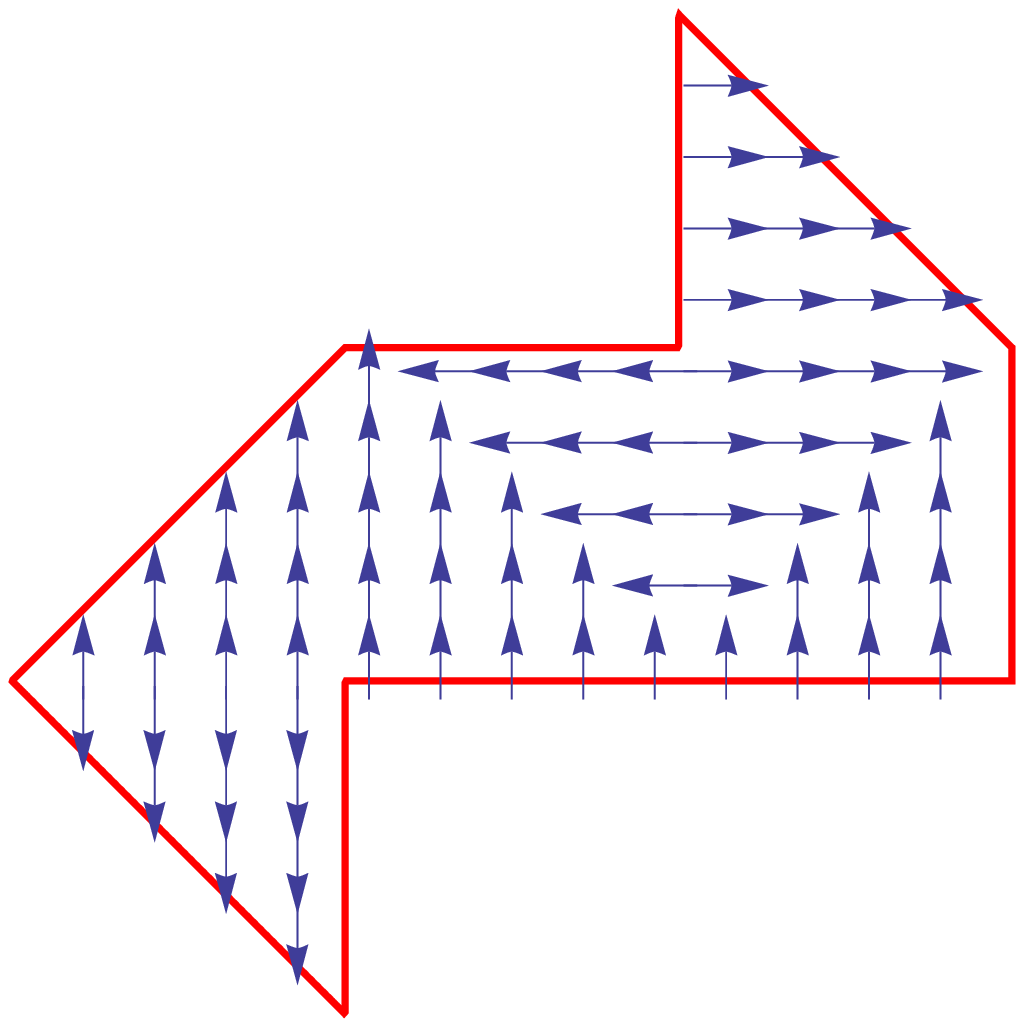} \\
\caption{(color online) Isospectral quantum billiards in an electric field. The arrows
indicate the direction of the constant electric field.}
\label{Fig_3}
\end{center}
\end{figure}

In this case the condition of isospectrality requires the potential $V(x,y)$ in each 
block to be the reflection of the potential on a neighboring block, along the dashed 
line separating the two. For instance $V(x,y)$ could be the potential generated by 
the interaction of an electron confined in any of the two domains of Fig.~\ref{Fig_2} 
with $7$ pointlike charges $q$ located at the center of mass of each building block. 
In Fig.~\ref{Fig_3} we display a simpler example of isospectral quantum billiard: 
the vector lines represent an electric field of constant magnitude $\mathcal{E}$ 
pointing in a given direction. Reversing the sign of $\mathcal{E}$ clearly corresponds
to inverting the directions of the vectors in the figure.

\textit{Numerical experiments---} 
It was noted by Wu, Sprung and Martorell in Ref.~\cite{WSM95} that the finite difference method
provides matrices which are exactly isospectral up to machine precision when applied to 
the calculation of the spectrum of the Laplacian over the two domains of Fig.~\ref{Fig_2} (clearly, 
the same grid  size is used in both cases). 

We have followed the same strategy of Ref.~\cite{WSM95} for the general problems discussed in
this paper, working with very fine grids (the maximum grid that we have generated contains 
$200521$ points) and using a collocation approach based on tent functions, which is equivalent
to a finite difference approach. 

In all our calculations we have found that the matrices obtained in the discretization of the 
problem are always {\sl exactly isospectral}, up to machine precision~\footnote{The isospectrality of
two domains may not be manifest when the images under reflection of a grid point in one building
block do not belong to the grid; in this case, the isospectrality is only obtained in the continuum
limit.} and therefore we will always report a single numerical value  for both domains.
This result provides a numerical confirmation of our previous results.

\begin{table}
\caption{\label{tab:table1} Lowest $10$ eigenvalues of
the inhomogeneous drums of Fig.~\ref{Fig_4}. Here $\Sigma_1=1$ and
$\Sigma_2=2$. }
\begin{ruledtabular}
\begin{tabular}{ccc}
n  & $E_n^{\rm (FD)}$ & $E_n^{\rm (EX)}$ \\
\hline
1  & 1.52189   & 1.51992  \\
2  & 2.63494   & 2.63002  \\
3  & 3.08334   & 3.07902  \\
4  & 4.58312   & 4.57697  \\
5  & 4.83882   & 4.83108  \\
6  & 6.23355   & 6.22662  \\
7  & 6.68975   & 6.67769  \\
8  & 7.71814   & 7.70122  \\
9  & 7.92551   & 7.91504  \\
10 & 8.66913   & 8.65750  \\
\end{tabular}
\end{ruledtabular}
\end{table}

The first example that we have studied is the inhomogeneous drum of Fig.~\ref{Fig_4}
with densities $\Sigma_1=1$ (lighther region) and $\Sigma_2=2$ (darker region). 
In Table \ref{tab:table1} we report the lowest $10$ eigenvalues of these domains
(the building blocks are triangles with angles $45^0$, $45^0$ and $90^0$ and shorter side
of length $\ell =2$): the second column contains the results obtained with finite difference with a 
grid containing $200521$ points; the third column contains the results obtained 
using Richardson extrapolation on a sequence of approximations obtained using 
grids with spacing $h = 1/4k$, with $k = 19, \dots, 30$.  For all the cases examined
this sequence has a monotonical behavior with decreasing $h$ and therefore the
extrapolation improves significantly the accuracy of the results. In the case of
a homogeneous membrane, where the very precise results of Ref.~\cite{Driscoll97} 
are available, this procedure applied on the sequence of eigenvalues obtained 
with the same grids used here allows one to obtain about four decimal correct 
for the lowest eigenvalues. We expect roughly the same accuracy here.

\begin{figure}
\begin{center}
\bigskip\bigskip\bigskip
\includegraphics[width=4cm]{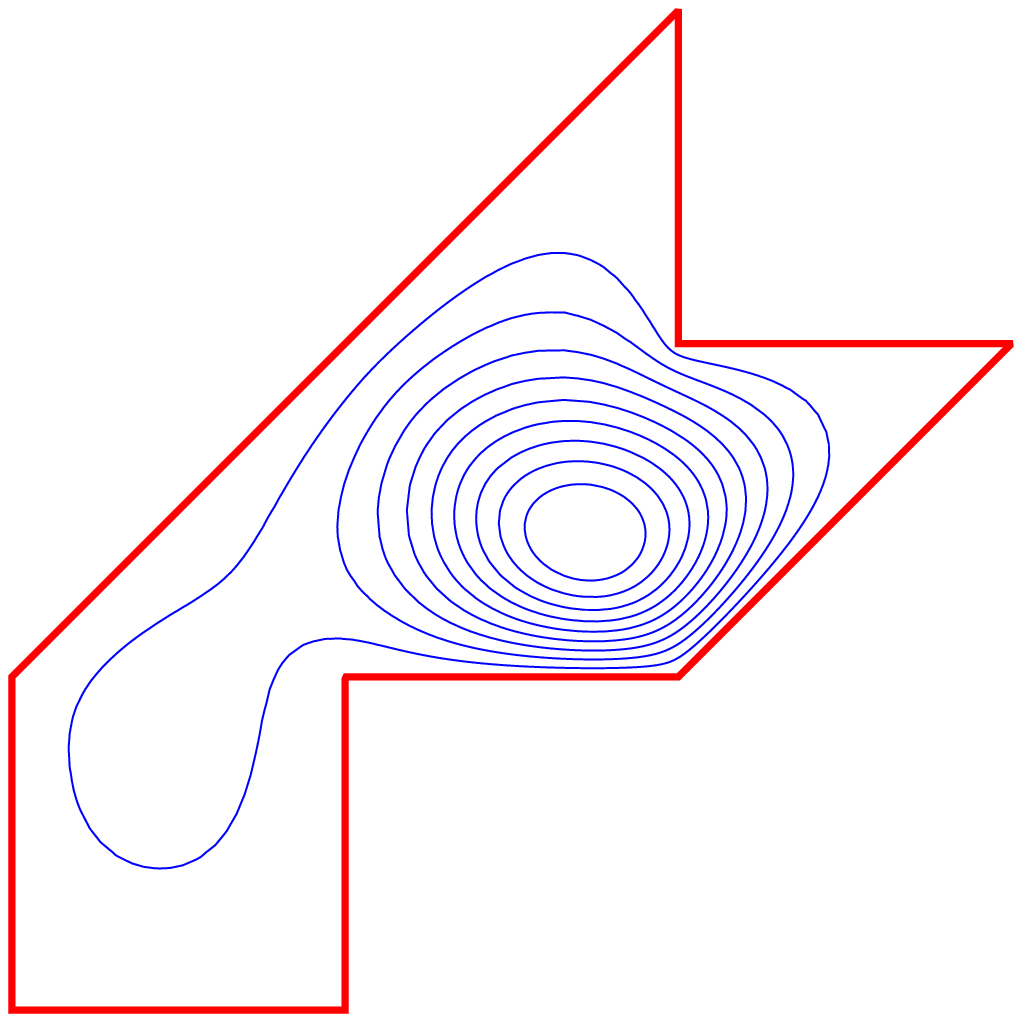} 
\includegraphics[width=4cm]{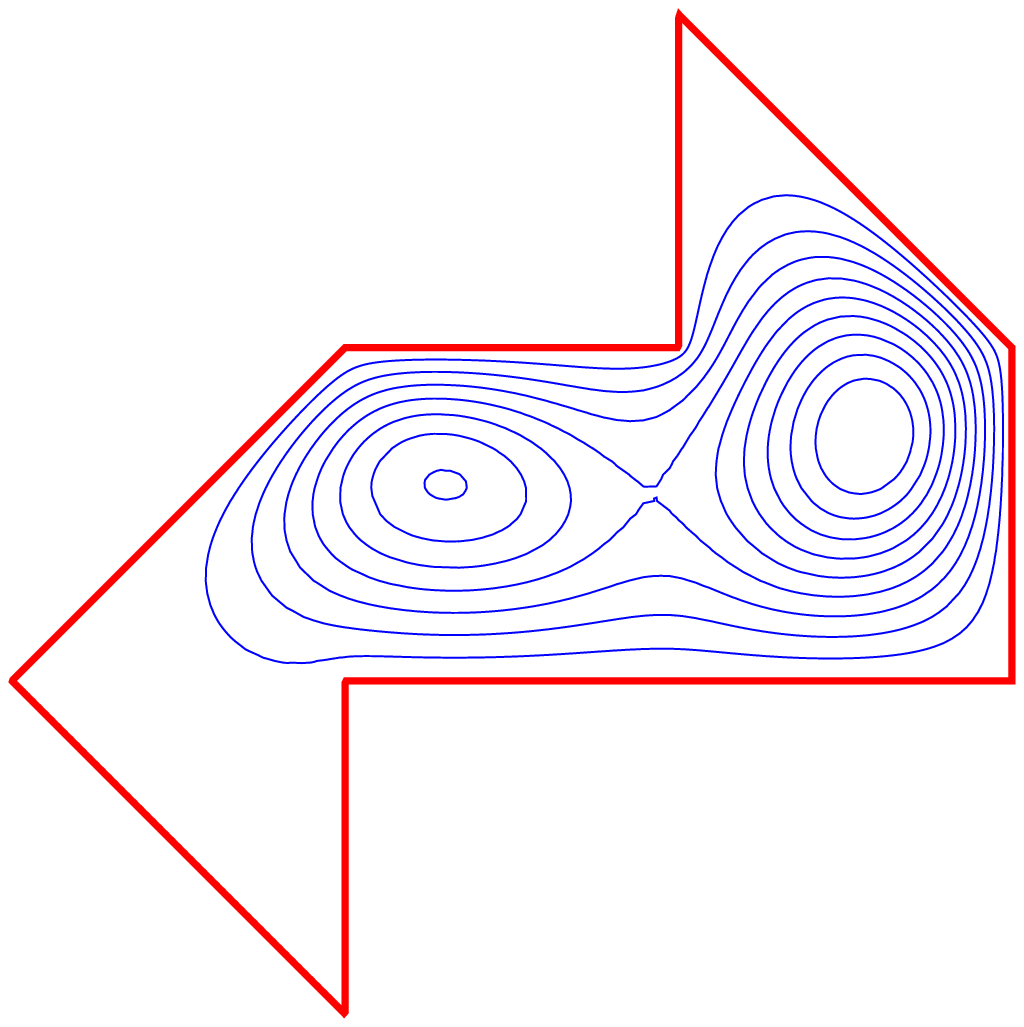} \\
\caption{(color online) Wave functions of the ground state of the 
isospectral Hamiltonians corresponding to Fig.\ref{Fig_3} with 
$\mathcal{E}=5$.}
\label{Fig_5}
\end{center}
\end{figure}

The second example that we consider is plotted in Fig.\ref{Fig_3}: 
in each building block an electric field of constant magnitude points at 
a given direction. Our numerical results have been obtained setting
$\hbar^2/2m=1$ and $e=1$, and considering an electric field
$\mathcal{E} = 5$. The wave functions for the ground state of this
problem in the two domains are plotted in Fig.~\ref{Fig_5}.
The corresponding eigenvalue obtained using the largest grid 
($200521$ points) is $E^{(FD)}_1 = -1.21302$; the value obtained
with extrapolation (see the discussion for the previous example) is $E^{(EX)}_1 = -1.21311$.

In conclusion, we have generalized the results of Gordon,Webb and
Wolpert \cite{GWW92} to a larger class of physical problems, which
include the case of inhomogeneous drums or of quantum billiards 
in an external field. We have proved that the domains found in 
Ref.~\cite{GWW92} are still isospectral when the density or the
potential in each building block is obtained from the reflection 
of the analogous quantities in the neighboring blocks, along the 
common border separating the two. In particular our results 
signal the possibility of building isospectral pairs of "ray-splitting" 
billiards, i.e. cavities with abrupt changes in the properties
of the medium filling it (see the original work by Couchman et 
al.~Ref.~\cite{Couchman92} and the works by Vaa et al., Ref.~\cite{Blumel03, Blumel05},
containing experimental verification of the theoretical semiclassical formulas). 

\begin{acknowledgments}
I thank dr Alfredo Aranda for reading the manuscript.
This research was supported by Sistema Nacional de Investigadores (M\'exico).
\end{acknowledgments}

\end{document}